\documentclass[twoside]{article}
\usepackage{fleqn,espcrc2}
\usepackage{graphicx}
\usepackage{here}

\newcommand{\AmS}{{\protect\the\textfont2
    A\kern-.1667em\lower.5ex\hbox{M}\kern-.125emS}}										
\def\beq{\begin{equation}}
\def\eeq{\end{equation}}
\def\bea{\begin{eqnarray}}
\def\eea{\end{eqnarray}}

\def\bq{\begin{quote}}
\def\eq{\end{quote}}

\def\nnb{\nonumber}
\def\ga{\left(}
\def\dr{\right)}

\def\lrar{\Longrightarrow}
																
\def\nnb{\nonumber}
\def\la{\langle}
\def\ra{\rangle}
\def\nin{\noindent}
\def\ba{\vspace*{-0.2cm}\begin{array}}
\def\ea{\end{array}\vspace*{-0.2cm}}

\def\b{$\bullet~$}
\def\als{\alpha_s}

\def\gg2{ \la\alpha_s G^2 \ra}
\def\gg3{g^3f_{abc}\la G^aG^bG^c \ra}
\def\ggg4{\la\als^2G^4\ra}

\title
{\bf{\boldmath
{\large $|V_{cd}|$ , $|V_{cs}|$ and $f_{D_{(s)}}$ from
 (semi) leptonic $D_{(s)}$-decays : signals of New Physics ? } }}

\author{Stephan Narison\thanks{Email: snarison@yahoo.fr}  \address {\footnotesize Laboratoire
de Physique Th\'eorique et Astroparticules, Universit\'e
de Montpellier II and CNRS, Case 070, Place Eug\`ene
Bataillon, 34095 - Montpellier Cedex 05,
France}
}
\begin{document}
\pagestyle{myheadings}
\markright{ }
\begin{abstract}
\noindent
We confront the recent improved measurements of the  $D_{(s)}$ (semi) leptonic decays 
with Lattice QCD (LQCD) with $n_f=3$ flavours and   QCD spectral sum rules (QSSR) predictions.  
$D\to\mu\nu_\mu $ leptonic width data compared with theoretical determinations of $f_D$, leads to the value of the CKM mixing angle : $|V_{cd}|= 0.230~(10)_{\rm exp}~(9)_{\rm th}$. 
Measured ratio of the  D semi-leptonic widths combined with LQCD and QSSR predictions leads to the average $|V_{cd}|/|V_{cs}|= 0.2175(88)$, and then to $|V_{cs}|= 1.068(47)$. We consider the previous determinations as improvements of the existing estimates.  Using the average data of the $D_s\to \mu\nu_\mu$ (resp. $D_s\to \tau\nu_\tau$) branching ratios,  one obtains : $|V_{cs}|f_{D_s}^{\mu}= (259\pm 12)~{\rm MeV}$ (resp) $|V_{cs}|f_{D_s}^{\tau}= (274\pm 13)~{\rm MeV}$, which  can be compared with the average of the Standard Model (SM) values from LQCD and QSSR $f_{D_s}=(240\pm 7)$ MeV. If one uses the present determination of  $ |V_{cs}|$, there is an agreement with the SM prediction within $1\sigma$. 
 If instead, we impose the unitarity constraint $ |V_{cs}|\leq 1$, or assume (as frequently done) $|V_{cs}|=|V_{ud}|$, we would obtain a deviation from the SM expectations ranging from 1.5 to 3 $\sigma$, therefore, signaling some New Physics beyond the SM. 
\end{abstract}
\maketitle
\section{Introduction}
 \nin
Good experimental and theoretical controls of the leptonic decay constants $f_P$ (analogue to $f_\pi$) of charmed $D_{(s)}$ and beautiful $B_{(s)}$ (pseudo)scalar mesons are of prime importance for understanding the dynamics of the heavy light quark systems
(overlap of the wavefunctions, heavy quark mass-behaviour,...) in QCD.  Many efforts have been devoted to this study both theoretically and experimentally \footnote{For a recent review see e.g. \cite{ROSNER}; For earlier reviews, see e.g. \cite{SNHL,SNB}.}. The leptonic partial width of the $D^+_q$ is normalized as:
\beq
\Gamma(D_q^+\to l^+\nu_l)={G_F^2\over 8\pi}|V_{cq}|^2 f^2_{D_q} m_l^2 M_{D_q}\ga 1-{m_l^2\over M_{D_q}^2}\dr^2
\eeq
where $q\equiv d,s$;  $M_{D_q}$ and $m_l$ are respectively  the masses of the $D^+_q$ meson and of the $l^+$ charged lepton. $G_F$ is
the Fermi constant and $|V_{cq}|$ is the CKM matrix element controlling the weak coupling of the $c$ and $q$ quarks. $f_{D_q}$ is the leptonic decay constant normalized like $f_\pi=(130.4\pm 0.2)$ MeV as:
\beq
(m_q+m_c)\la 0|\bar q\gamma_5 c |D_q\ra =f_{D_q}M^2_{D_q}~.
\eeq
The previous leptonic
process is helicity suppressed ($\sim m_l^2$) and the last factor in parenthesis is the familiar phase space factor. While the electron mode is tiny due to strong helicity suppression, the measurement of the $\tau$ mode involves the detection of  additionnal neutrinos, such that, the muon mode is experimentally the cleanest and most accessible one. Therefore, we shall mainly consider the $D^+_q\to \mu^+\nu_\mu$ data which (in principle) provide the most precise measurements, and show for a comparison the one from $D_s \to \tau\nu_\tau$ decay. 

\section{Determination of $|V_{cd}|$}
\nin
For determining $f_D$ from $D\to \mu\nu_\mu$ decay, it has become usual to assume that $|V_{cd}|=|V_{us}|$, which is an extra input in the analysis. In addition, the extraction of $|V_{us}|$ from $Kl_3$ is affected by the uncertainties on the value of the K-form factor, while $\tau$-decay data give a slightly different (though consistent) value \cite{PICH}. 
 \vspace*{-0.5cm}
{\scriptsize
\begin{table*}[H]
\setlength{\tabcolsep}{0.5pc}
\caption{\scriptsize $f_{D_{(s)}}$ and $f_{D_s}/f_D$ from LQCD with $n_f=3$ flavours  and QSSR.}
\label{tab:fd}
\begin{tabular}{llll}
&\\
\hline
\hline
\\
Method &$f_D$&$f_{D_s}$&$f_{D_s}/f_D$\\
\\
\hline
\hline
\\
 LQCD\cite{LQCD0}& $201\pm 17.3$& $249\pm 16.3$&$1.24\pm 0.07$ \\
 LQCD\cite{LQCD1}& $208\pm 4$& $241\pm 3$&$1.162\pm 0.009$ \\
 \\
QSSR&&&\\
Full QCD \cite{SNHL,SNB,SNRATIO}& $203\pm 20$& $235\pm 24$&$1.15\pm 0.04$ \\
Full QCD \cite{PENIN}&$195\pm 20$&&\\
Analytic Cont.\cite{BORDES}&&&$1.16\pm 0.03$\\

&\\
\hline
&\\
Average  &$202\pm 8.3$&$241.7\pm 9.7$&$1.178\pm 0.022$ \\
&\\
\hline
\hline
\end{tabular}
\end{table*}
}
\nin
The good agreeement between the measured and theoretical values of $f_D$ \cite{ROSNER,CLEOa,STONE}, allows us to perform (with a good confidence) an opposite, though natural, procedure,  by relying on the theoretical value of $f_D$ for extracting more accurately $|V_{cd}|$. 
By combining the most recent measured  $D\to \mu\nu_\mu$ branching ratio from CLEO  \cite{CLEOa,STONE}:
\beq
Br(D^+\to \mu^+\nu_\mu)= (3.82\pm 0.32\pm 0.09)\times 10^{-4}~,
\eeq
with the averaged value from LQCD and QSSR \`a la SVZ \cite{SVZ} given in Table \ref{tab:fd} \footnote{We have only considered the ratio obtained in \cite{BORDES} as the decay constants obtained there are systematically lower than other predictions, which may question the systematics of the Analytic Continuation method \cite{SNB}, that can cancel in the ratio.  
We have only quoted the most recent 3-loop predictions relevant for our discussions. More complete references  are given in the reviews \cite{SNHL,SNB}. 
 We have not taken the weighted average which should be dominated by the most accurate LQCD determination where the validity of the errors are still under discussions \cite{MORTE}. Instead, we have taken a na\"\i ve average of the central value and have averaged the errors quadratically.}, 
 and using the $D^+$ lifetime of $(1040\pm 7)$ ps \cite{PDG},
we deduce:
\beq
|V_{cd}|= 0.230\pm 0.010_{\rm exp}  \pm 0.009_{\rm th}~,
\label{eq:vcd}
\eeq
to be compared with :
\beq
|V_{us}|=0.2165\pm 0.0026_{\rm exp}\pm 0.0005_{\rm th}
\eeq
 from $\tau$-decay \cite{PICH}. 
\section{Determination of  $|V_{cs}|$}
\nin
Semileptonic decays of the light $K$ and heavy $B$ mesons have given most of the CKM mixing parameters.  In the case of the process $D^0\to P_q^- l^+\nu_l$, ($P_d^-\equiv \pi^- $ and $P_s\equiv K^-$) and ($l\equiv e,~\mu$), where one can neglect the ratio $(m_l/m_c)^2$, the differential rate can be expressed as:
\beq
{d\Gamma\over dq^2}\Big{\vert}_{D^0\to P_q^- l^+\nu_l}={G_F^2\over 24\pi^3}p_q^3|V_{cq}|^2|f^{D\to P}_+(q^2)|^2~,
\eeq
where $q^2$ is the invariant mass squared of the $ l\nu$ system; $f^{D\to P}_+(q^2)$ is the form factor and $p_q$ is the momentum
of the light meson.  CLEO data lead to the result \cite{CLEO1}:
\beq
{|f_+^{D\to\pi}(0)||V_{cd}|\over |f_+^{D\to K}(0)||V_{cs}|}= 0.188~(8)~(2)~,
\eeq
 where the errors are respectively statistical and systematical. In order to use this data, we update
the value of the ratio of form factors from QSSR \cite{SNFORM} by using the recent value $\bar m_s(2~\rm GeV)=(96.3\pm 17.5)$ MeV \cite{SNmass} instead of the one $m_s(2~\rm GeV)=115.7$ MeV used
in the original paper. We have also used the value of  $m_c$ in \cite{SNHL,SNB}~\footnote {There is an unfortunate misprint in \cite{SNFORM}. Instead of 0.007 for the error in Eq. (24), one should read 0.07.}.
Then, we obtain: 
  \beq
r_D\equiv { |f_+^{D\to K}(0)|/ |f_+^{D\to\pi}(0)|} = 1.11\pm 0.07~,
 \eeq
 which  has almost the same value as the previous one. The error has been multiplied by $\sqrt{2}$ in order to take into account unknown higher order terms in the QCD series.
 \vspace*{-0.5cm}
{\scriptsize \begin{table*}[H]
\setlength{\tabcolsep}{1.2pc}
\caption{\scriptsize Theoretical predictions of form factors.}
\label{tab: formfactor}
\begin{tabular}{ll}
&\\
\hline
\hline
\\
Methods&$r_D\equiv { |f_+^{D\to K}(0)|/|f_+^{D\to\pi}(0)|}$ \\
\\
\hline
\hline
\\
LQCD \cite{LQCD0}& $1.149\pm 0.040\pm 0.119$\\
LCSR\cite{BALL}& $1.19\pm 0.06$\\
QSSR\cite{SNFORM} (updated)& $1.11\pm 0.07$\\
&\\
\hline
&\\
Average  &$1.155\pm 0.043 $ \\

&\\
\hline
\hline
\end{tabular}
\end{table*}
}
\nin
which we compare in Table \ref{tab: formfactor} with some other determinations, from which we deduce the average:
 \beq
r_D\equiv { |f_+^{D\to K}(0)|/ |f_+^{D\to\pi}(0)|} = 1.155\pm 0.043~.
 \eeq
Some remarks are in order here:  \\
\b The determinations of the ratio compared to the absolute values are always more accurate in the QSSR calculations due to the cancellations of different systematics. The sensitivity to the higher state contributions (continuum threshold-dependence), to PT radiative corrections (which tend to cancel in the ratio), to the exact value of the charm quark mass (absent to leading order) ,... are less pronounced in the direct evaluation of the ratio than in each individual form factor. The dependence on the sum rule variable  of the prediction is also weaker as shown in the figure of \cite{SNFORM}.  \\
\b This value does not come from a numerical fit but from a semi-analytic expression where all different sources of corrections can be easily controlled.\\
\b The sum rule used here has successfully predicted (though with a modest accuracy) the absolute value of the form factor $f^{B\to \pi}_+(0)= (0.23\pm 0.02)$ for $B\to \pi\mu\nu_\mu$ and of other semileptonic decays \cite{SNSEMI} and the decay rate $B\to K^*\gamma$~\cite{SNBK}.\\
\b As a good illustration of this property, the LQCD recent most precise prediction of the ratio \cite{LQCD1}:
\beq
f_{D_s}/f_D= 1.162\pm 0.009~,
\label{eq:fdsfdlatt}
\eeq
agrees within a digit with the central value from QSSR prediction obtained one decade earlier \cite{SNRATIO,SNHL,SNB} \footnote{An alternative estimate using the analytic continuation method  leads to similar result but with a smaller error \cite{BORDES} as given in Table \ref{tab:fd}.}:
\beq
f_{D_s}/f_D= 1.15\pm 0.04~,
\label{eq:fdsfd}
\eeq
which is a strong indication on the reliability of the QSSR approach despite its modest accuracy \footnote{As indicated in  \cite{SNRATIO,SNHL,SNB}, the  error in this ratio has been enlarged by $\sqrt{2}$  for a crude estimate of the unknown $m_s^4\alpha_s^2$ corrections, which I wish will be available in the near future. However, after inspection of the analytic expression of the spectral function this crude estimate can give an overestimate of the true error.},  and a compatibility of the LQCD and QSSR results as often encountered in different channels. 
The same remark also holds for the absolute value of the decay constants. \\
\b It is also important to notice that the previous ratio in Eq. (\ref{eq:fdsfd}) comes from a semi-analytic  functional dependence, which is easy to control rather than from a numerical fit.  \\
Using the previous inputs and taking into account previous remarks, we can deduce with a quite good accuracy :
\beq
{|V_{cd}|\over |V_{cs}|}= 0.2171~(95)_{\rm exp}~(81)_{\rm th}~.
\label{eq:ratiockm}
\eeq
which is compared in Table \ref{tab:vcd} with some alternative determinations \cite{CLEO1}.
Using the previous value of $|V_{cd}|$ into Eq. (\ref{eq:ratiockm}), one can deduce:
\beq
|V_{cs}|= 1.059~(65)_{\rm exp}~  (58)_{\rm th}.
\label{eq:vcs}
\eeq
Considering as a final result, the (weighted) average of the three determinations in Table \ref{tab:vcd}, we deduce:
\beq
{|V_{cd}|\over |V_{cs}|}= 0.2175~(88)~,~~~~~~|V_{cs}|= 1.068~(47)~,
\label{eq:ckmfinal}
\eeq
to be compared with the usual assumption:
\beq
|V_{cs}|\simeq |V_{ud}|= 0.97377\pm 0.00027.
\label{eq:vud}
\eeq
We consider these results as improvements of earlier results obtained in \cite{SNFORM}. 
Further tests of the assumption $|V_{cs}|\simeq |V_{ud}|$ and some eventual deviations from the unitarity conditions can be reached in future improved measurements of $D$-decays. These eventual deviations may also reveal some New Physics beyond the SM expectations.
 \vspace*{-0.5cm}
{\scriptsize
\begin{table*}[H]
\setlength{\tabcolsep}{0.08pc}
\caption{\scriptsize $|V_{cd}|/|V_{cs}|$\&$|V_{cs}|$ using CLEO data + Theoretical mehods.\\
The 1st (resp) 2nd errors are experimental (resp) theoretical. \\
The 3rd error in CLEO
is due to $f^{D\to P}_+(0)$ from LQCD.}
\label{tab:vcd}
\begin{tabular}{llll}
&\\
\hline
\hline
\\
Method &$|V_{cd}|/|V_{cs}|$&$|V_{cs}|$& Comments\\
\\
\hline
\hline
\\
 LQCD\cite{LQCD1}& $0.2257(209) (20)$& $1.070(70)(10)$& \\
CLEO\cite{CLEO1}& $0.2138(100) (30)(110)$& $1.075(69)(47)(55)$&LQCD\\
This work&$0.2171(95) (81)$& $1.059(65)(58)$&QSSR\\
&\\
\hline
&\\
Average  &$0.2175(88)$&1.068(47) \\
&\\
\hline
\hline
\end{tabular}

\end{table*}
}
\nin
 \vspace*{-0.5cm}
{\scriptsize \begin{table*}[H]
\setlength{\tabcolsep}{0.75pc}
\caption{\scriptsize$D_s\to\mu\nu_\mu$ branching ratios.}
\label{tab: datads}
\begin{tabular}{lll}
&\\
\hline
\hline
\\
Exp.&$B_{\mu} \times 10^3$&$B_{\phi\pi}(\%)$ \\
\\
\hline
\hline
\\
CLEO-c\cite{CLEO-c}& $5.94\pm 0.66\pm 0.31$&\\
BELLE\cite{BELLE}& $6.44\pm 0.76\pm 0.52$&\\
CLEO\cite{CLEO}&$6.2\pm 0.8\pm 1.3\pm 1.6$&$3.6\pm 0.9$\\
BEATRICE\cite{BEATRICE}&$8.3\pm 2.3\pm 0.6\pm 2.1$& $3.6\pm 0.9$\\
ALEPH\cite{ALEPH}&$6.8\pm 1.1\pm 1.8$&$3.6\pm 0.9$\\
BABAR\cite{BABAR}&$6.74\pm 0.83\pm 0.26\pm 0.66$&$4.71\pm 0.46$\\
&\\
\hline
&\\
Average  &$6.13\pm 0.57^{+)} $& \\
(no rad. corr.)&&\\
&$6.33\pm 0.47$&\\

&\\
\hline
\hline
\end{tabular}
{\footnotesize 
\begin{quote}
$^{+)}$ Exclude $\phi\pi^+$ mode normalizations.
\end{quote}}
\end{table*}
}
\nin
\section{The value of $f_{D_s}$ from $D_s \to \mu\nu_\mu$}
\nin
We shall use the previous value of the CKM angle $|V_{cs}|$ in Eq. (\ref{eq:ckmfinal}) \footnote{In the current literature, one
often assumes $|V_{cs}|=|V_{ud}|$.} for extracting the 
value of $f_{D_s}$ from $D_s\to \mu\nu_\mu$, which has been emphasized to be the
cleanest experimental mode. We shall use the average of the different data in Table \ref{tab: datads}
and the $D_s$ lifetime of $(0.500\pm 0.004)$ ps \cite{PDG}. One can notice that the average including or excluding the $\phi\pi^+$
modes is almost the same. However, taking into account that the normalization to the $\phi\pi^+$ modes
induce more systematic uncertainties \cite{ROSNER}, we shall only consider the 1st average which does not
use this normalization. Radiative corrections will decrease the branching ratio by about 2\%, which we shall include
in the extraction of the decay constant. Then, we deduce:
\beq
f_{D_s}^\mu= (242.5\pm 10.7_{\rm V_{cs}}\pm 11.5_{\rm exp})~{\rm MeV}~.
\label{eq:fds1}
\eeq
The uncertainty is comparable with the one from the theoretical average given in Table \ref{tab:fd}, and is larger than the one quoted in \cite{ROSNER,STONE}. The main reason is that the result of \cite{ROSNER,STONE} uses the
assumption $|V_{cs}|=|V_{ud}|$ (see Eq. (\ref{eq:vud}), where  $|V_{ud}|$ has a tiny error.The central value of $f_{D_s}^{\rm exp}$ is also smaller than in \cite{ROSNER,STONE} because the one of $|V_{cs}|$ obtained in Eq. (\ref{eq:ckmfinal}) is larger than that of $|V_{ud}|$.  If instead, we only impose the unitarity constraint $ |V_{cs}|\leq 1$, one can deduce the lower bound: 
\beq
|V_{cs}|\leq 1\lrar f_{D_s}^\mu\geq (259\pm 12)~{\rm MeV}~~~~(90\%~{\rm  CL})~.
\label{eq:fds2}
\eeq
The usual assumption $|V_{cs}|=|V_{ud}|$ used in \cite{ROSNER,STONE} would imply:
\beq
|V_{cs}|=|V_{ud}|\lrar f_{D_s}^\mu= (266\pm 12)~{\rm MeV}~.
\label{eq:fds3}
\eeq

\nin
\section{The value of $f_{D_s}$ from $D_s \to \tau\nu_\tau$}
\nin
Though expected to be less accurately measured than the $D_s \to \mu\nu_\mu$ decay, it is informative to repeat the previous analysis for the $D_s \to \tau\nu_\tau$ data, which are given in Table \ref{tab: datatau}. 
 \vspace*{-0.5cm}
{\scriptsize \begin{table*}[H]
\setlength{\tabcolsep}{2.5pc}
\caption{\scriptsize$D_s\to\tau\nu_\tau$ branching ratios.}
\label{tab: datatau}
\begin{tabular}{ll}
&\\
\hline
\hline
\\
Exp.&$B_{\tau} \times 10^2$ \\
\\
\hline
\hline
\\
CLEO-c\cite{CLEO-c}& $8.0\pm 1.3\pm 0.4$\\
CLEO-c\cite{CLEO-c2}& $6.17\pm 0.71\pm 0.36$\\
ALEPH\cite{ALEPH}& $5.8\pm 0.8\pm 1.8$\\
L3\cite{L3}&$7.4\pm 2.8\pm 1.6\pm 1.8$\\
OPAL\cite{OPAL}&$7.0\pm 2.1\pm 2.0$\\
&\\
\hline
&\\
Average  &$6.57\pm 0.63 $ \\

&\\
\hline
\hline
\end{tabular}
\end{table*}
}
\nin
From the averaged branching ratio, one can deduce in MeV \footnote{As mentioned in \cite{ROSNER}, the $\tau$ rate does not need to be radiatively corrected.}:
\bea
f_{D_s}^{\rm \tau}&=& 256.6\pm 11.3_{\rm V_{cs}}\pm 12.4_{\rm exp}~:~|V_{cs}|~ {\rm in~{Eq.}~(\ref{eq:ckmfinal}})
 \nnb\\
&\geq& 273.5\pm 13.2_{\rm exp}~:~|V_{cs}|\leq 1 ~~(90\%~{\rm  CL})\nnb\\
&=& 280.9\pm 13.5_{\rm exp}~:~|V_{cs}|=|V_{ud}|~,
\label{eq:tau}
\eea
where one can notice that the value of $f_{D_s}$ from $\tau$-data is about 1$\sigma$ higher than the one from $\mu$-data:
\beq
f_{D_s}^{\rm \tau}/f_{D_s}^\mu= 1.06\pm 0.07~.
\label{eq:ratiofdsfd}
\eeq
\section{Comparison with theoretical predictions}
\nin
One can compare the previous experimental values with the theoretical predictions reviewed in \cite{ROSNER,SNHL,SNB}, which we have selected in Table \ref{tab:fd}. From this table, the average of the direct determinations of $f_{D_s}$ is:
\beq
f_{D_s}^{\rm dir}= (241.7\pm 9.7)~{\rm MeV}~,
\label{eq:fdsth1}
\eeq
while the one obtained from the average value of $f_D$ multiplied by the ratio $f_{D_s}/f_D$ is:
\beq
f_{D_s}^{\rm ratio}=(238.0\pm 11.8)~{\rm MeV}~,
\label{eq:fdsth2}
\eeq
from which we deduce the na\"\i ve average :
\beq
f_{D_s}^{\rm th}=(240\pm 7)~{\rm MeV}~,
\label{eq:fdsth}
\eeq
where the errors have been added quadratically. We consider this result as the final theoretical result to be compared with experiments.\\
\b {\it $f_{D_s}$ from $D_s\to \mu\nu_\mu$-decay:} one can notice that the value of $f_{D_s}$ obtained in Eq. (\ref{eq:fds1}) by using the value of $|V_{cs}|$ in Eq. (\ref{eq:ckmfinal}) agrees quite well with the previous average of the SM predictions in Eq. (\ref{eq:fdsth}).  Discrepancies between the experimental and theoretical numbers start to be visible ($\geq 1.5~ \sigma$) when using the unitarity constraint for $|V_{cs}|\leq 1$ [ see Eq. (\ref{eq:fds2}) ] and becomes $2~\sigma$ when one uses the additional assumption  $|V_{cs}|= |V_{ud}|$ [ see Eq. (\ref{eq:fds3}) ] .  \\
\b {\it $f_{D_s}$ from $D_s\to \tau\nu_\tau$-decay:} here, a comparison of the theoretical prediction shows larger discrepancies ranging from 1 $\sigma$ in the case of $|V_{cs}|$ from Eq. (\ref{eq:ckmfinal}) to 3 $\sigma$ in the case  $|V_{cs}|= |V_{ud}|$. 
\section{Interpretations: signals of New Physics ?}
\nin
\b Some deviations from the SM expectations can be manifest if the $|V_{cs}|$ satisfies the unitarity constraints. Therefore, more precise determinations of $|V_{cs}|$ and to a lesser extent of $|V_{cd}|$ are required for a sharp
test of the SM predictions. Some eventual deviations of these CKM matrix elements from unitarity constraints can also signal
some departures from the SM expectations independently of the $f_{D_s}$ values.\\
\b We have shown that the deviation of  $f_{D_s}$ [Eqs. (\ref{eq:fds1}) to (\ref{eq:tau})] from the SM expectations [Eq.(\ref{eq:fdsth}) can be indeed quite large ($3\sigma$ in the $\tau$ channel) but a sharp conclusion needs a better control of the CKM mixing matrix  $|V_{cs}|$. \\
\b We have also found in Eq. (\ref{eq:ratiofdsfd}) that the value of $f_{D_s}$ from $D_s\to \tau\nu_\tau$ decay can be larger than the one from $D_s\to \mu\nu_\mu$ decay. More precise measurements of  $f_{D_s}$ in both channels can then provide an universality test of the $W\mu\nu_\mu$ and $W\tau\nu_\tau$ vertices for these processes. At present, $\tau$ decay data show a slight violation \cite{PICH}:
\beq
B_{W\to \tau\nu_\tau}/B_{W\to \mu\nu_\mu}= 1.039 \pm 0.013~,
\eeq
which can permit some new interactions (not necessary the same) at these vertices. \\
\b Using the fact that the value of the decay constant $f_{D_s}$ deviates from the SM predictions, some attempts to explain this deviation  as due to non-standard effective interactions or/and to new scalar particles or/and to leptoquarks have been discussed in the literature \cite{KRONFELD}.  Analysis of different models beyond the SM are beyond the scope of this paper  though planned to be done in a future work.
\section{Conclusions}
\nin
We have re-examined the consequences of the recent data on (semi)leptonic $D_{(s)}$ decays by confronting them with the LQCD and QSSR theoretical calculations. \\
\b First, these data have been used to estimate  the CKM mixing angles $|V_{cd}|$ and $|V_{cs}|$. The results are given in Eq. (\ref{eq:vcd}) and in Eqs.  (\ref{eq:ratiockm}) to (\ref{eq:ckmfinal}). \\
\b The resulting value of $|V_{cs}|$ have
been used together with the $D_s \to \mu\nu_\mu$ and $D_s\to \tau\nu_\tau$ data for extracting the leptonic decay constant $f_{D_s}$, which behaves like $1/|V_{cs}|$ [ Eqs.  (\ref{eq:fds1})  to  (\ref{eq:tau})]. \\
\b Comparing the  experimental value of $f_{D_s}$ with the theoretical averaged LQCD and QSSR determinations in Eq. (\ref{eq:fdsth}), 
we conclude that a sharp detection of an eventual deviation of the SM predictions from these processes, needs either better determinations of the CKM mixing angles $|V_{cd}|$ and $|V_{cs}|$ than the one obtained in this paper, or a strict validity of the frequently used assumption $|V_{cs}|=|V_{ud}|$ or of the unitarity condition $|V_{cs}|\leq 1$. In these cases, the data indicate  deviations from the SM expectations from 1.5 to 3 $\sigma$, therefore, signaling some New Physics beyond the SM.
\section*{Acknowledgements}
\nin
It is a pleasure to thank Sheldon Stone and Helmut Vogel for communicating the different CLEO data and for
some comments. This work has been motivated by the talk given by Helmut Vogel on the behalf of the CLEO collaboration at the 14th International Conference QCD 08 (7-12th July 2008, Montpellier-FR).

\end{document}